\documentclass[aps, superscriptaddress, twocolumn]{revtex4-2}
\usepackage{amsfonts}
\usepackage{amssymb}
\usepackage{amsmath}
\usepackage{graphics}
\usepackage{graphicx}
\usepackage{subfigure}
\usepackage{dcolumn}

\usepackage{bm}
\usepackage{color}
\usepackage{epstopdf}

\definecolor{Gray}{gray}{0.1}

\DeclareMathAlphabet\mathbfcal{OMS}{cmsy}{b}{n}

\begin{document}

\title{Disorder-induced effects in high-harmonic generation process in
fullerene molecules}
\author{H.K. Avetissian}
\affiliation{Centre of Strong Fields Physics at Research Institute of Physics, Yerevan State University,
Yerevan 0025, Armenia}
\author{S. Sukiasyan}
\affiliation{Centre of Strong Fields Physics at Research Institute of Physics, Yerevan State University,
Yerevan 0025, Armenia}
\affiliation{Max-Planck-Institut f\"ur Kernphysik, Saupfercheckweg 1, 69117 Heidelberg, Germany}
\author{H.H. Matevosyan}
\affiliation{Institute of Radiophysics and Electronics NAS RA, Ashtarak
0203, Armenia}
\author{G.F. Mkrtchian}
\thanks{Email: mkrtchian@ysu.am}
\affiliation{Centre of Strong Fields Physics at Research Institute of Physics, Yerevan State University,
Yerevan 0025, Armenia}

\begin{abstract}
{
The objective of this article is to investigate the profound nonlinear
optical response exhibited by inversion symmetric fullerene molecules under
the influence of different types of disorders described by the Anderson
model. Our aim is to elucidate the localization effects on the spectra of
high harmonic generation in such molecules. We show that the
disorder-induced effects are imprinted onto molecules' high-harmonic
spectrum. Specifically, we observe a presence of strong even-order harmonic
signals already for relatively small disorders. The odd-order harmonics
intrinsic for disorder-free systems are generally robust to minor disorders.
Both diagonal and off-diagonal disorders lift the degeneracy of states,
opening up new channels for interband transitions, leading to the
enhancement of the high-harmonic emission. The second harmonic signal has a
special behavior depending on the disorder strength. Specifically in the
case of diagonal disorder, the second harmonic intensity exhibits a
quadratic scaling with the disorder strength, which enables the usage of the
harmonic spectrum as a tool in measuring the type and the strength of a
disorder.}

\end{abstract}

\maketitle

\section{Introduction}

High harmonic generation (HHG) is a nonlinear key process\cite%
{corkum1993plasma} in strong-field physics, which has made great
achievements over the past decades in the gaseous phase of matter. Since the
discovery of HHG in solids \cite%
{ghimire2011observation,zaks2012experimental,schubert2014sub,vampa2015linking,luu2015extreme,ghimire2019high}
and novel nanostructures \cite{geim2013van} of fascinating properties, much
theoretical (for review see \cite{avetissian2022efficient} and references
therein) and experimental \cite{Yoshikawa,hafez2018extremely,le2018high}
attention has been paid to the multiphoton and HHG processes in novel
nanostructures. The interest in the HHG process is huge. Particularly, it
gives an access to the frequency range that is difficult to achieve in other
ways \cite{avetissian2015relativistic}, and thanks to the extremely large
coherent bandwidth of harmonic spectra HHG enables spectroscopy in
attosecond resolution \cite{krausz2009attosecond}. According to the HHG
spectra in crystals, one can observe the dynamical Bloch oscillations \cite%
{luu2015extreme}, Peierls \cite{bauer2018high}, and Mott \cite{silva2018high}
transitions. The understanding of the high-harmonic spectrum can help in
reconstruction of the electronic \cite{vampa2015all} and topological
properties of the materials \cite%
{luu2018measurement,avetissian2020high,avetissian2022high}. The generation
of harmonics has recently been reported in liquids \cite{luu2018extreme} and
in amorphous solids \cite{you2017high}, which ensures that the periodicity
is not a necessary condition for HHG in condensed matter systems. Some
theoretical works found that liquids \cite%
{zeng2020impact,xia2022role,ding2023high}, doped materials \cite%
{huang2017high,almalki2018high,yu2019high,yu2019enhanced}, solutions \cite%
{xia2022theoretical}, and disordered semiconductors \cite%
{orlando2018high,orlando2019macroscopic,chinzei2020disorder} can produce
harmonics efficiently.

The disorder is always present in crystals, and it is of particular interest
to investigate its influence on the HHG process in novel nanostructures.
Since the fundamental works by Anderson \cite{anderson1958absence} and
co-workers \cite{abrahams1979scaling}, it is known that the presence of the
disorder in a lattice structure is a key factor defining the extension of
the electronic wave function. Depending on the dimensionality of the system
Anderson localization changes its character \cite{evers2008anderson}.
Particularly, even for small disorder strengths, all states in a disordered
system with dimension below 2 are localized in a small fraction of the
lattice. It is well known that the HHG process is very sensitive to
electronic wave function extension \cite{lewenstein1994theory} and therefore
high harmonic spectroscopy can reveal disorder-induced Anderson
localization. Latter has been shown for finite linear chain \cite%
{pattanayak2021high}. Anderson localization is a fundamental wave phenomenon
and takes place also for finite nanostructured materials with a sufficiently
large number of sites. For systems with a finite number of atoms, particular
interest represent the planar quantum dots based on graphene or the case of
fullerenes with lattice topologies of nodes distributed over a closed
surface. The most known fullerene is C$_{60}$ \cite{kroto1985c60}, the
discovery of which triggered the study of many other carbon nanostructures 
\cite{iijima1991helical,geim2013van}. Today, graphene quantum dots \cite%
{gucclu2014graphene} and fullerene molecules \cite{fowler2007atlas} are
promising materials for extreme nonlinear optics. Theoretically strong HHG
emission is predicted in C$_{60}$ and C$_{70}$ molecules \cite%
{zhang2005optical,zhang2006ellipticity,zhang2020high,avetissian2021high}.
The analysis in mentioned works is for disorder-free systems and it is
unclear how the disorder leaves its mark on the HHG and sub-cycle electronic
response. In recent years the effect of the disorder, impurities and
vacancies on HHG in solid materials has been studied in the papers \cite%
{orlando2018high,yu2019enhanced,yu2019high,pattanayak2020influence,iravani2020effects,chinzei2020disorder,hansen2022doping,xia2022theoretical,orlando2022ellipticity}%
. These investigations reveal situations where an imperfect lattice might
actually amplify HHG in comparison to a perfectly ordered lattice \cite%
{yu2019enhanced,yu2019high,pattanayak2020influence,hansen2022doping}. In
finite systems, like fullerene molecules, our current understanding
indicates that the effect of the disorder on the electromagnetic response is
predominantly confined to the linear regime of interaction \cite%
{harigaya1994optical}.

In the present work, we develop a microscopic theory for the nonlinear
interaction of the fullerene molecules with the strong electromagnetic
radiation of linear polarization taking into account diagonal and
off-diagonal disorders. We also consider the electron-electron interaction
(EEI) in the Hartree-Fock approximation \cite{avetissian2021high}. In
particular, we consider C$_{60}$ and C$_{180}$ molecules with the same
icosahedron point group symmetry, but with different number of atoms, which
allows one to study the size effect as well. By means of the dynamical
Hartree-Fock approximation, we study the dependence of the HHG spectrum on
the disorder type and strength. Our concept can be easily generalized to
other molecules of this family.

This paper is organized as follows. In Sec. II, the model and the basic
equations are formulated. In Sec. III, we represent the main results.
Finally, conclusions are given in Sec. IV.

\section{The model and theoretical methods}

Our target represents a carbon based nanostructure, with sites distributed
over a closed surface, interacting with mid-infrared laser pulse. In
particular, we consider the lattice topology of fullerene buckyballs C$_{60}$
and C$_{180}$. Both molecules are invariant under the inversion with respect
to the center of mass and have icosahedral point group ($I_{h}$) symmetry.
We assume neutral fullerene molecules, described within the tight-binding
theory where the interball hopping is much smaller than the on-ball hopping,
and the EEI is described within the extended Hubbard approximation \cite%
{chiappe2015can}. The total Hamiltonian reads:\textrm{\ }

\begin{equation}
\widehat{H}=\widehat{H}_{\mathrm{TB}}+\widehat{H}_{\mathrm{C}}+\widehat{H}_{%
\mathrm{int}},  \label{1H}
\end{equation}%
where%
\begin{equation}
\widehat{H}_{\mathrm{TB}}=\sum_{i\sigma }\epsilon _{i}c_{i\sigma }^{\dagger
}c_{j\sigma }-\sum_{\left\langle i,j\right\rangle \sigma }\left(
t_{ij}+\beta _{ij}\right) c_{i\sigma }^{\dagger }c_{j\sigma }  \label{2H}
\end{equation}%
is the free, tight-binding, fullerene Hamiltonian. Here $c_{i\sigma
}^{\dagger }$\ ($c_{i\sigma }$) creates (annihilates) an electron with the
spin polarization $\sigma =\left\{ \uparrow ,\downarrow \right\} $\ at the
site $i$\ ($\overline{\sigma }$\ is the opposite to $\sigma $\ spin
polarization), $\epsilon _{i}$ is the energy level at the site $i$, and $%
\left\langle i,j\right\rangle $\ runs over all the first nearest-neighbor
hopping sites with the hopping integral $\left( t_{ij}+\beta _{ij}\right) $
between the sites $i$ and $j$. In the absence of disorder, $\epsilon _{i}=0$%
, $\beta _{ij}=0$, 
and the hopping integral between the nearest-neighbor atoms at$\ $positions $%
\mathbf{r}_{i}$ and $\mathbf{r}_{j}$ is approximated by $t_{ij}=t_{0}+\alpha
\left( d_{0}-\left\vert \mathbf{r}_{i}-\mathbf{r}_{j}\right\vert \right) $ 
\cite{zhang2005optical}. Taking $d_{0}=1.54\mathrm{\mathring{A}}$ and by
fitting the energy gap for C$_{60}$, we have determined the average hopping
constant $t_{0}=2.17\ \mathrm{eV}$ and the electron-lattice coupling
constant $\alpha =3.52\ \mathrm{eV/\mathring{A}}$. Input coordinates for the
C$_{60}$ and C$_{180}$ are generated with the program Fullerene via a
face-spiral algorithm \cite{fowler2007atlas}. The initial structure is
further optimized by a force field specifically designed for fullerenes \cite%
{schwerdtfeger2013program}. In the real-space tight-binding Hamiltonian (\ref%
{2H}) the disorder is introduced with randomly distributed site energies $%
\epsilon _{i}$, called as diagonal disorder, or with random hopping
integrals $\beta _{ij}$ called as off-diagonal disorder. We assume for the
random variables $\epsilon _{i}$ and $\beta _{ij}$ to have probability
distributions $P\left( \epsilon _{i},V_{\mathrm{on}}\right) $ and $P\left(
\beta _{ij},V_{\mathrm{off}}\right) $, where 
\begin{equation}
P\left( x,\Delta \right) =\left\{ 
\begin{array}{c}
\frac{1}{2\Delta },\ -\Delta \leq x\leq \Delta \\ 
0,\ \mathrm{otherwise}%
\end{array}%
\right. .  \label{dist}
\end{equation}%
Here the quantities $V_{\mathrm{on}}$ and $V_{\mathrm{off}}$ are the
distribution widths describing the strength of the disorders. Similar to
other macroscopic or mesoscopic systems, fullerene molecules are subject to
different types of disorders \cite{fischer1993order}. Diagonal disorder
arises from impurities or the randomness of the surrounding medium \cite%
{tummala2015static, d2016charges}, while off-diagonal disorder emerges from
fluctuations in bond lengths around their average value caused by lattice
distortions \cite{harigaya1994optical}. The assessment of disorder strength
for both types of disorders suggests that a value $V_{\mathrm{on}\text{,}%
\mathrm{off}}\sim 0.1t_{0}$ would be a reasonable estimate.

The second term in the total Hamiltonian (\ref{1H}) describes the extended
Hubbard Hamiltonian with the $U$ and $V$ terms included. Within the
Hartree-Fock approximation, the Hamiltonian $\widehat{H}_{\mathrm{C}}$ is
approximated by%
\begin{equation}
\begin{aligned} &\widehat{H}_{C}^{HF}=U\sum_{i}\left(
\overline{n}_{i\uparrow }-\overline{n}_{0i\uparrow }\right) n_{i\downarrow }
\\ &+U\sum_{i}\left( \overline{n}_{i\downarrow }-\overline{n}_{0i\downarrow
}\right) n_{i\uparrow }+\sum_{\left\langle i,j\right\rangle }V_{ij}\left(
\overline{n}_{j}-\overline{n}_{0j}\right) n_{i} \\ &-\sum_{\left\langle
i,j\right\rangle \sigma }V_{ij}c_{i\sigma }^{\dagger }c_{j\sigma }\left(
\rho_{ji}^{\left( \sigma \right)} -\left\langle c_{i\sigma }^{\dagger
}c_{j\sigma }\right\rangle _{0}\right) , \label{2s} \end{aligned}
\end{equation}%
with on- and inter-site Coulomb repulsion energies $U$\ and $V_{ij}$,
respectively, where $n_{i\sigma }=c_{i\sigma }^{\dagger }c_{i\sigma }$ is
the density operator and $n_{i}=n_{i\uparrow }+n_{i\downarrow }$ is the
total electron density for the site $i$. Here $\overline{n}_{i\sigma
}=\left\langle c_{i\sigma }^{\dagger }c_{i\sigma }\right\rangle $ and $\rho
_{ji}^{\left( \sigma \right) }=\left\langle c_{i\sigma }^{\dagger
}c_{j\sigma }\right\rangle $. Since the distance $d_{ij}$\ between the
nearest-neighbor pairs varies over the system, we scale inter-site Coulomb
repulsion as: $V_{ij}=Vd_{\min }/d_{ij}$, where $d_{\min }$\ is the minimal
nearest-neighbor distance. For all calculations we use the ratio $V=0.4U$\ 
\cite{martin1993coulomb}. In this representation the initial density matrix $%
\rho _{0ji}^{\left( \sigma \right) }=\left\langle c_{i\sigma }^{\dagger
}c_{j\sigma }\right\rangle _{0}$ is calculated with respect to the
tight-binding Hamiltonian $\widehat{H}_{\mathrm{TB}}$, neglecting the EEI
Hamiltonian, i.e. $\widehat{H}_{C}^{HF}=0$.

The last term in the total Hamiltonian (\ref{1H}) is the light-matter
interaction part that is described in the length-gauge:%
\begin{equation}
\widehat{H}_{\mathrm{int}}=e\sum_{i\sigma }\mathbf{r}_{i}\cdot \mathbf{E}%
\left( t\right) c_{i\sigma }^{\dagger }c_{i\sigma },  \label{intH}
\end{equation}%
where $\mathbf{r}_{i}$ is the position vector and $\mathbf{E}\left( t\right)
=f\left( t\right) E_{0}{\hat{\mathbf{e}}}\cos \omega t$ represents the
electric field, with 
the frequency $\omega $, amplitude $E_{0}$, polarization vector $\hat{%
\mathbf{e}}$ and pulse envelope $f\left( t\right) =\sin ^{2}\left( \pi t/%
\mathcal{T}\right) $, where $\mathcal{T}$ is the pulse duration. From the
Heisenberg equation we obtain evolutionary equations for the single-particle
density matrix $\rho _{ij}^{\left( \sigma \right) }=\left\langle c_{j\sigma
}^{\dagger }c_{i\sigma }\right\rangle $: 
\begin{equation}
\begin{aligned} i\hbar \frac{\partial \rho _{ij}^{\left( \sigma \right)
}}{\partial t}&=\sum_{k}\left( \tau _{kj\sigma }\rho _{ik}^{\left( \sigma
\right) }-\tau _{ik\sigma }\rho _{kj}^{\left( \sigma \right) }\right)
+\left( V_{i\sigma }-V_{j\sigma }\right) \rho _{ij}^{\left( \sigma \right) }
\\ &+e\mathbf{E}\left( t\right) \left( \mathbf{r}_{i}-\mathbf{r}_{j}\right)
\rho _{ij}^{\left( \sigma \right) }-i\hbar \gamma \left( \rho _{ij}^{\left(
\sigma \right) }-\rho _{0ij}^{\left( \sigma \right) }\right) , \label{evEqs}
\end{aligned}
\end{equation}%
where 
\begin{equation}
V_{i\sigma }=\sum_{j\alpha }V_{ij}\left( \rho _{jj}^{\left( \alpha \right)
}-\rho _{0jj}^{\left( \alpha \right) }\right) +U\left( \rho _{ii}^{\left( 
\overline{\sigma }\right) }-\rho _{0ii}^{\left( \overline{\sigma }\right)
}\right)  \label{Vij}
\end{equation}%
\ and 
\begin{equation}
\tau _{ij\sigma }=-\epsilon _{i}\delta _{ij}+t_{ij}+\beta _{ij}+V_{ij}\left(
\rho _{ji}^{\left( \sigma \right) }-\rho _{0ji}^{\left( \sigma \right)
}\right)  \label{tauij}
\end{equation}%
are defined via$\ $the$\ $density matrix $\rho _{ij}^{\left( \sigma \right)
} $\ and its initial value. In addition, we assume that the system relaxes
at a rate $\gamma $ to the equilibrium $\rho _{0ij}^{\left( \sigma \right) }$
distribution. While the evolutionary equations (\ref{evEqs}) for the
single-particle density matrix in the HF approximation offer a reasonably
accurate approximation for describing light absorption effects in fullerene
molecules even for larger Coulomb interaction energies $U\sim 2t_{0}-4t_{0}$%
\ \cite{harigaya1994optical}, it is important to mention that systems with
the large $U$ require a treatment beyond the HF approximation \cite%
{lev2014dynamics}, such as the Kadanoff-Baym-Keldysh approach \cite%
{kadanoff1962quantum,keldysh1965diagram}, to properly account for Coulomb
interactions.

The incorporation of disorders into the evolutionary equations, as
demonstrated by Eqs. (\ref{evEqs}), (\ref{Vij}), and (\ref{tauij}), occurs
through dual mechanisms. Firstly, by means of the initial single-particle
density matrix $\rho _{0ij}^{\left( \sigma \right) }$, which is calculated
in relation to the tight-binding Hamiltonian (\ref{2H}). Secondly, through
the adjustment of the mean-field hopping integrals (\ref{tauij}). In the
subsequent analysis, we will explore scenarios in which only one type of
disorder is present. In the context of diagonal disorder, we establish $%
\beta _{ij}$\ as zero, indicating the modification of solely on-site
energies. In the case of off-diagonal disorder, $\epsilon _{i}$\ is set to
zero, signifying the alteration of exclusively hopping integrals. For all
results we repeat the calculations with equally distributed variables $%
\epsilon _{i}$\ or $\beta _{ij}$\ according to distribution (\ref{dist}) and
then take the ensemble average.

\begin{figure}[tbp]
\includegraphics[width=0.48\textwidth]{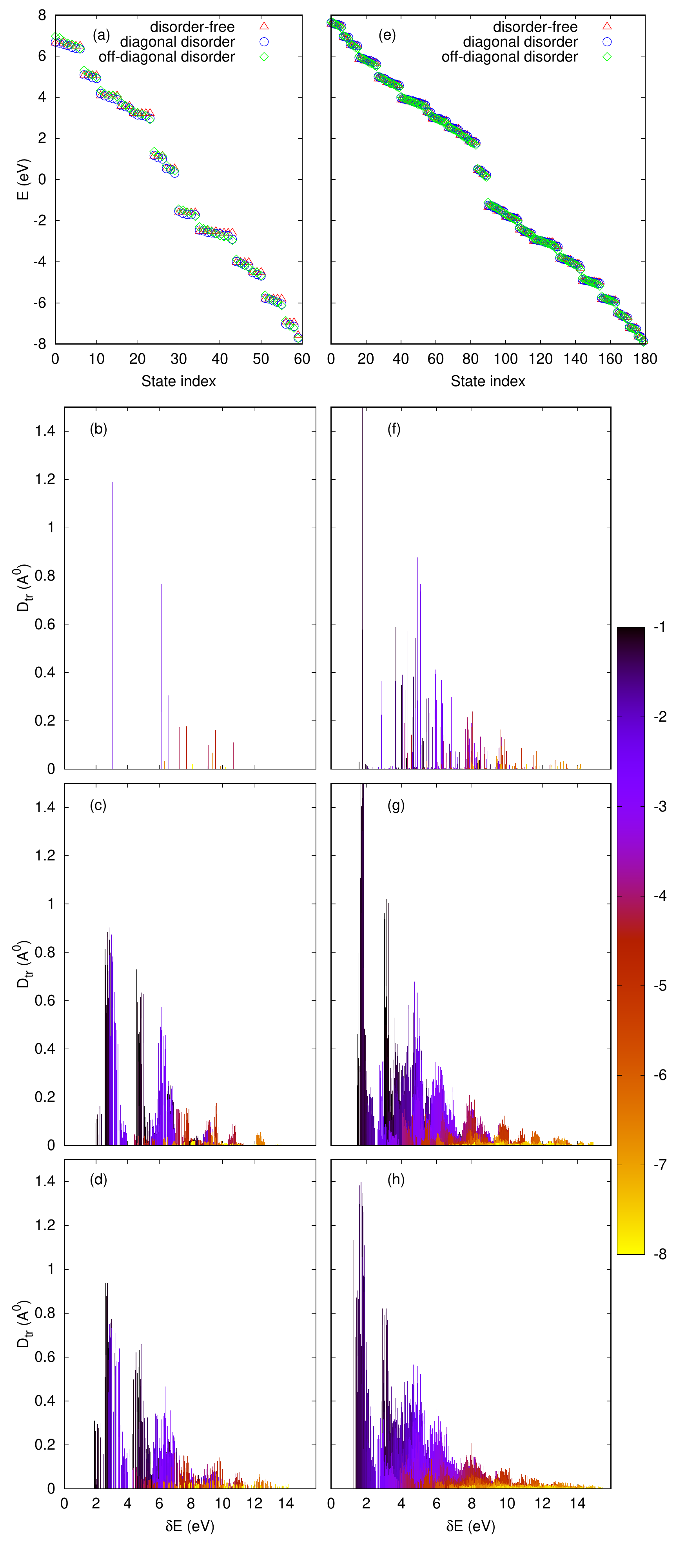}
\caption{Eigenenergies of disorder-free and disordered fullerene molecules
are displayed for $C_{60}$ (a) and $C_{180}$ (e). The absolute values of
matrix elements of the transition dipole moment for interband transitions
are shown for $C_{60}$ (b, c, d) and for $C_{180}$ (f, g, h). Panels (b) and
(f) correspond to the case without disorders ($V_{\mathrm{on}}=V_{\mathrm{off%
}}=0)$. Panels (c) and (g) represent the scenario with only diagonal
disorder ($V_{\mathrm{on}}=0.5\ \mathrm{eV}$, $V_{\mathrm{off}}=0$). Panels
(d) and (h) concern the case of solely off-diagonal disorder ($V_{\mathrm{on}%
}=0$, $V_{\mathrm{off}}=0.5\ \mathrm{eV}$). The color bar within the figure
provides the energy range (in eV) of the valence band.}
\label{fig1}
\end{figure}
\begin{figure}[tbp]
\includegraphics[width=0.48\textwidth]{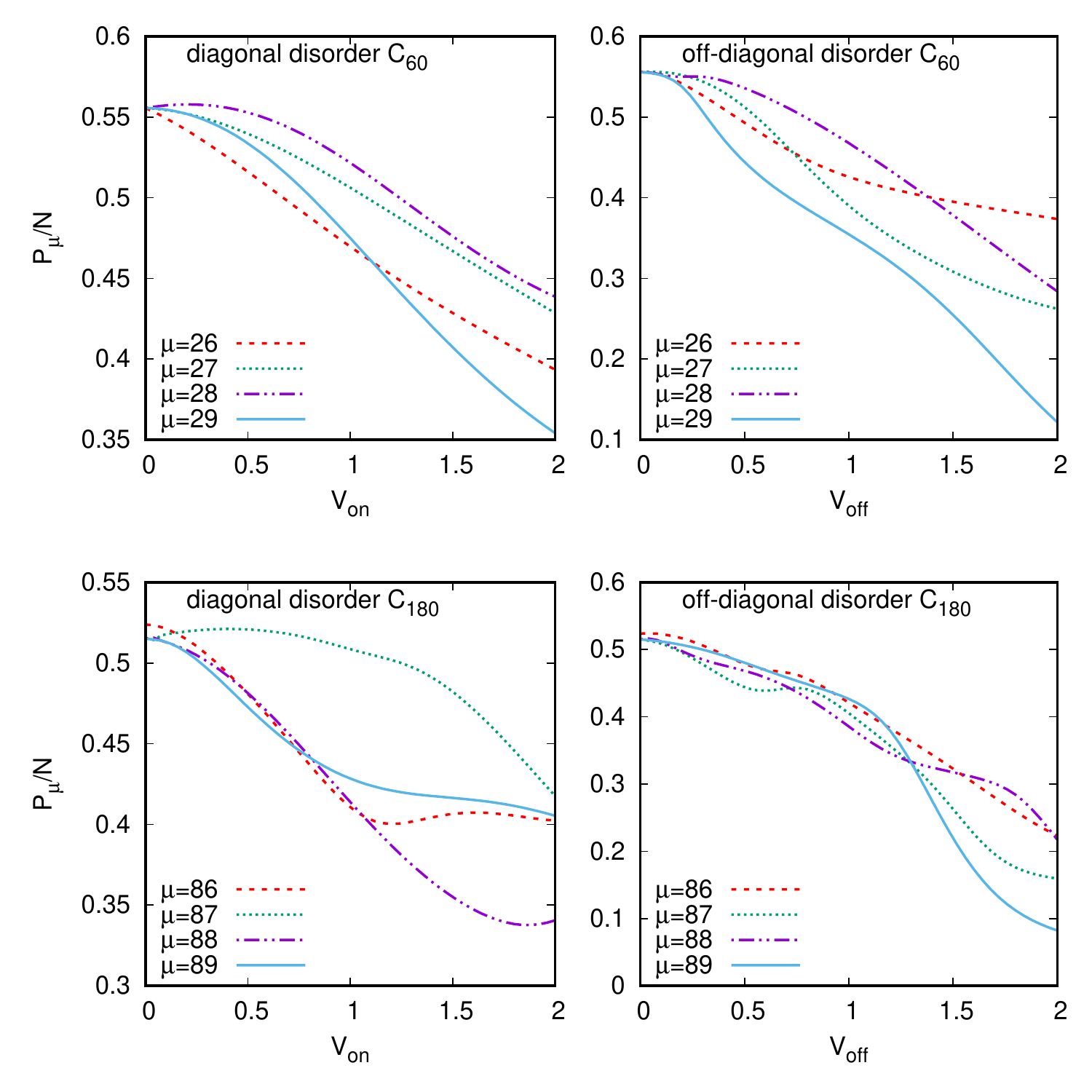}
\caption{The normalized inverse participation number for four states near
the Fermi level as a function of disorder strength for C$_{60}$ and C$_{180}$%
.}
\label{fig2n}
\end{figure}
\begin{figure*}[tbp]
\includegraphics[width=0.85\textwidth]{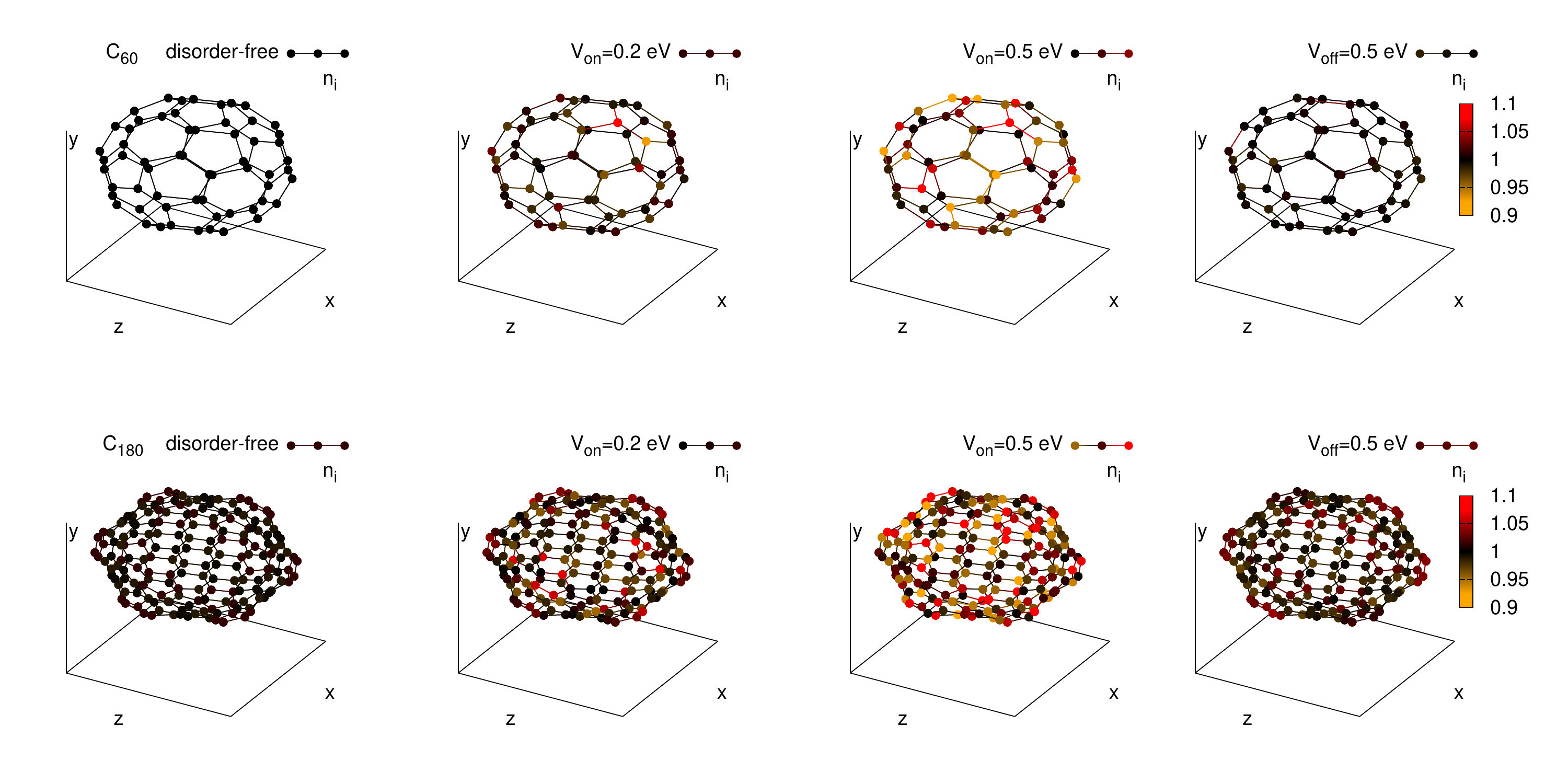}
\caption{The equilibrium occupations $n_{i}$ of sites in the 3D color-mapped
molecular structures for $C_{60}$ (upper panel) and for $C_{180}$ (lower
panel). The first configuration is disorder free. Next 3 configurations are
for diagonal disorder, $V_{\mathrm{on}}=0.2$ and $0.5$ eV, and for
off-diagonal disorder $V_{\mathrm{off}}=0.5$ eV.}
\label{fig2}
\end{figure*}

\section{Results}

\subsection{\label{subsec:states}Eigenstates localization, transition
channels and eigenenergies}

First we consider the influence of the disorder on the eigenstates and
eigenenergies of considered systems prior the interaction with the strong
laser pulse. These results are obtained by numerical diagonalization of the
tight-binding Hamiltonian (\ref{2H}) including the diagonal and off-diagonal
disorders. Numerically diagonalizing the tight-binding Hamiltonian $\widehat{%
H}_{\mathrm{TB}}$ we find the eigenstates $\psi _{\mu }\left( i\right) $,
eigenenergies $\varepsilon _{\mu }$ ($\mu =0,1..N-1$), and dipole transition
matrix elements $\mathbf{d}_{\mu ^{\prime }\mu }=e\sum_{i}\psi _{\mu
^{\prime }}^{\ast }\left( i\right) \mathbf{r}_{i}\psi _{\mu }\left( i\right) 
$. The results are shown in Fig. \ref{fig1} for C$_{60}$ and C$_{180}$.
Figures \ref{fig1} (a) and (e) show the eigenspectrum of fullerene molecules
without and with disorders included. The differences in the eigenspectrum
for the three (off, diagonal and off-diagonal) disorder cases are almost
indistinguishable. Compared with the C$_{60}$ molecule, the C$_{180}$ has
more degenerated states. In Figs. \ref{fig1} (b-d,f-h) the absolute value of
transition dipole moment versus energy difference is displayed, which shows
that both, the diagonal and off-diagonal, disorders lift the degeneracy of
states and break the inversion symmetry opening up new channels for
interband transitions. Besides, we observe that the first dipole-allowed
transition gaps are narrowed. As we will see, these two factors have
considerable influences on the HHG spectrum. 

In contrast to the eigenspectrum, the eigenstates of the system become
partially localized due to the Anderson mechanism. To provide a quantitative
characterization of localization in the $\mu $-th eigenstates, we also
calculate the inverse participation number%
\begin{equation}
P_{\mu }=\left( \sum_{i=0}^{N-1}\left\vert \psi _{\mu }\left( i\right)
\right\vert ^{4}\right) ^{-1},  \label{IPN}
\end{equation}%
which provides a measure of the fraction of sites over which the wavepacket
is spread \cite{kramer1993localization}. In the case of delocalized states
evenly distributed over the entire system, the inverse participation ratio
becomes equal to $N$, while for a state located in a single site, $P_{\mu
}=1 $. For the finite systems under consideration, where the thermodynamic
limit is not applicable, the inverse participation number serves as an
indicator of the localization tendency. Additionally, we will explore
disorders with strengths comparable to or smaller than the nearest neighbor
hopping matrix element, $t_{0}$. Although the localization arises from the
Anderson mechanism, it should be noted that for moderate disorder strengths,
we will have partial localization. To achieve complete localization,
disorders with strengths significantly larger than the hopping matrix
element would be required, which is not experimentally feasible for the
considered systems. As illustrated in Fig. \ref{fig2n}, we plot the
normalized inverse participation number $P_{\mu }/N$\ for four states near
the Fermi level as a function of disorder strength for C$_{60}$\ and C$%
_{180} $ molecules. From Fig. \ref{fig2n}, it becomes evident that there is
a clear trend towards localization in both cases. The localization effect is
weak and more pronounced in the case of off-diagonal disorder. We further
investigate the impact of localized eigenstates on the charge distribution
within the fully-filled valence band. To visualize this, we construct the
initial density matrix $\rho _{0ij}^{\left( \sigma \right) }$ by populating
electron states in the valence band according to the zero temperature
Fermi-Dirac distribution $\rho _{0ij}^{\left( \sigma \right) }=\sum_{\mu
=N/2}^{N-1}\psi _{\mu }^{\ast }\left( j\right) \psi _{\mu }\left( i\right) $%
. Fig. \ref{fig2} displays the equilibrium site occupations, $n_{i}=\rho
_{0ii}^{\left( \uparrow \right) }+\rho _{0ii}^{\left( \downarrow \right) }$,
for C$_{60}$ (upper panel) and C$_{180}$ (lower panel) respectively. We
observe a significant change in the initial configuration when diagonal
disorder is introduced, deviating from the initially homogeneous
distribution. However, in the presence of off-diagonal disorder, despite the
localization of individual states, the initial configuration remains
relatively close to a homogeneous distribution.

\subsection{Disorder-induced effects in HHG process}

To study the HHG process in $C_{60}$ and $C_{180}$ fullerene molecules we
evaluate the high-harmonic spectrum by Fourier Transformation of the dipole
acceleration, $\mathbf{a}\left( t\right) =d^{2}\mathbf{d(t)}/dt^{2}$, where
the dipole momentum is defined as $\mathbf{d}\left( t\right) =e\sum_{i\sigma
}\mathbf{r}_{i}\rho _{ii}^{\left( \sigma \right) }\left( t\right) $: 
\begin{equation}
\mathbf{a}\left( \Omega \right) =\int_{0}^{\mathcal{T}}\mathbf{a}\left(
t\right) e^{i\Omega t}W\left( t\right) dt,  \label{aW}
\end{equation}

\begin{figure}[tbp]
\includegraphics[width=0.45\textwidth]{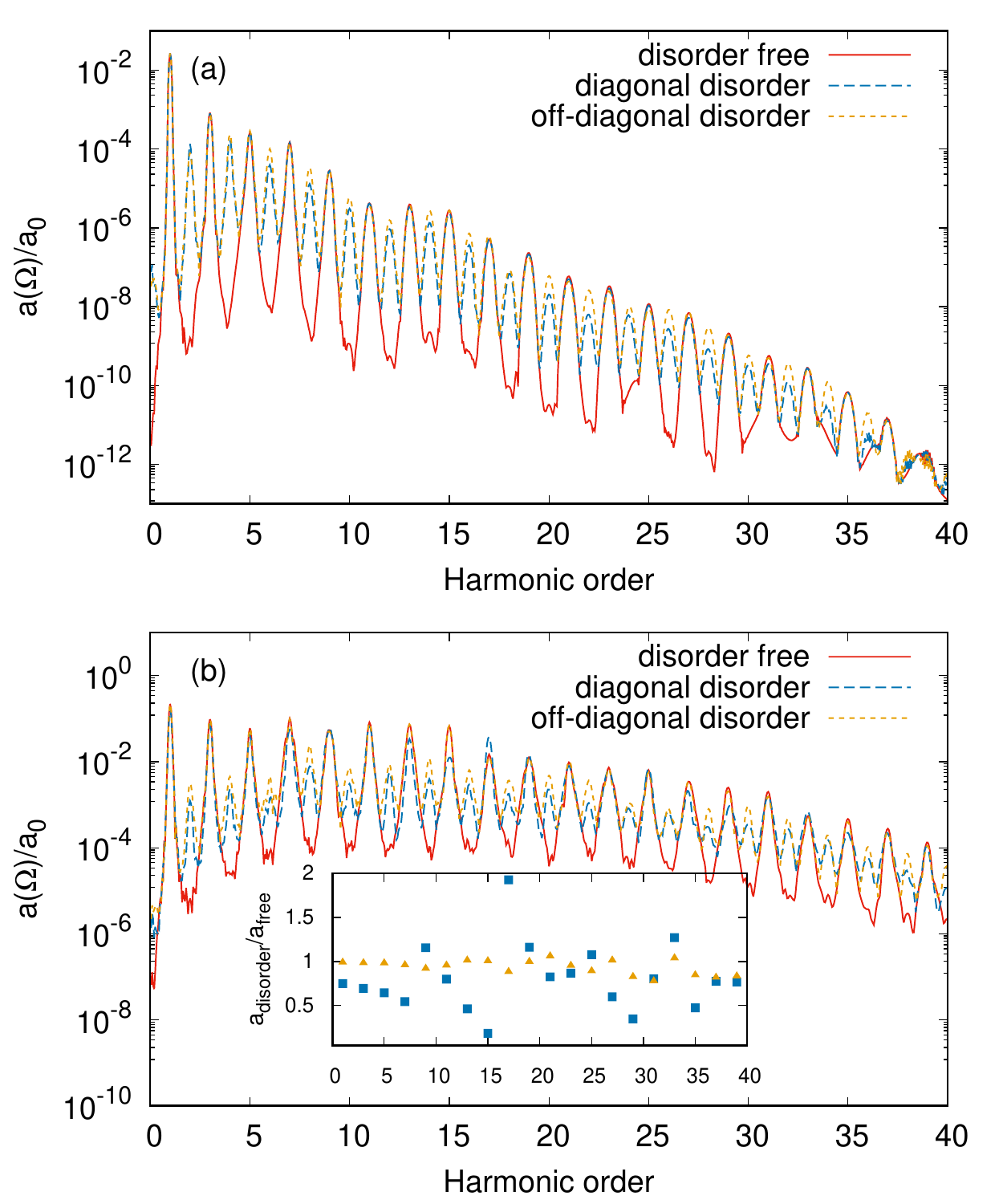}
\caption{Normalized HHG spectra, $a\left( \Omega \right) /a_{0}$, in
logarithmic scale for $C_{60}$ (a) and for $C_{180}$ (b). For disorders we
have taken $V_{\mathrm{on}}=0.1\ \mathrm{eV}$ and $V_{\mathrm{off}}=0.1\ 
\mathrm{eV}$. The wave amplitude is taken to be $E_{0}=0.2\ \mathrm{V/%
\mathring{A}}$. The inset shows the ratio of the spectra with and without
disorders for odd harmonics.}
\label{fig3}
\end{figure}
where $W\left( t\right) $ is the window function to suppress small
fluctuations \cite{zhang2018generating} and to decrease the overall
background (noise level) of the harmonic signal. As a window function we
take the pulse envelope $f\left( t\right) $. We assume the excitation
frequency $\omega =0.2\ \mathrm{eV}/\hbar $ is much smaller than the typical
gap $\sim 2\ \mathrm{eV}$.

The linearly polarized laser pulse is assumed to have $20$ wave cycles: $%
T=40\pi /\omega $. To obtain the mean picture which does not depend on the
orientation of the molecule with respect to laser polarization, we take the
wave polarization vector as $\hat{\mathbf{e}}=\left\{ 1,1,1\right\} $. The
Coulomb repulsion energy due to the screening can be considered to be in the
range $1-3.0$\ $\mathrm{eV}$ \cite{martin1993coulomb,zhang2003hartree}. The
spectra are calculated for moderate Coulomb repulsion energy, $U=2\ \mathrm{%
eV}$. While reported relaxation rates for fullerene molecules typically are
around $0.01t_{0}$\ \cite{harigaya1994optical} for linear optical effects,
we adopt a value of $\hbar \gamma =0.1\ \mathrm{eV}$\ in this study. This
choice is motivated by the presence of disorder and high-intensity laser
excitation, which are expected to enhance the relaxation rate. It is worth
noting that the effects investigated in this paper exhibit only a minor
dependence on the relaxation rate. For the convenience, we normalize the
dipole acceleration by the factor $a_{0}=e\overline{\omega }^{2}\overline{d}%
, $ where $\overline{\omega }=1\ \mathrm{eV}/\hbar $ and $\overline{d}=1\ 
\mathrm{\mathring{A}}$. The power radiated at the given frequency is
proportional to $\left\vert \mathbf{a}\left( \Omega \right) \right\vert ^{2}$%
. The time propagation of Eq. (\ref{evEqs}) is performed by the eight-order
Runge-Kutta algorithm. The time step is taken to be $\Delta t=2\times
10^{-2}\ \mathrm{fs}$.

Fig. \ref{fig3} shows the high-harmonic spectra for $C_{60}$ and $C_{180}$
systems in strong laser field. In considered systems for the disorder-free
case the inversion symmetry leads to the presence of only odd harmonics in
the HHG spectrum. As we mention from Fig. \ref{fig3}, even for small
disorder strengths we observe even harmonic signals comparable to odd ones.
Due to the size effects even harmonics are more pronounced in $C_{60}$
molecule. The appearance of even-order harmonics is connected with the
localization of eigenstates which breaks the inversion symmetry. The odd
harmonics, as can be seen in the inset of Fig. \ref{fig3}(b), demonstrate
relative stable behavior in the presence of the disorder, especially for the
off-diagonal one.

\begin{figure}[tbp]
\includegraphics[width=0.39\textwidth]{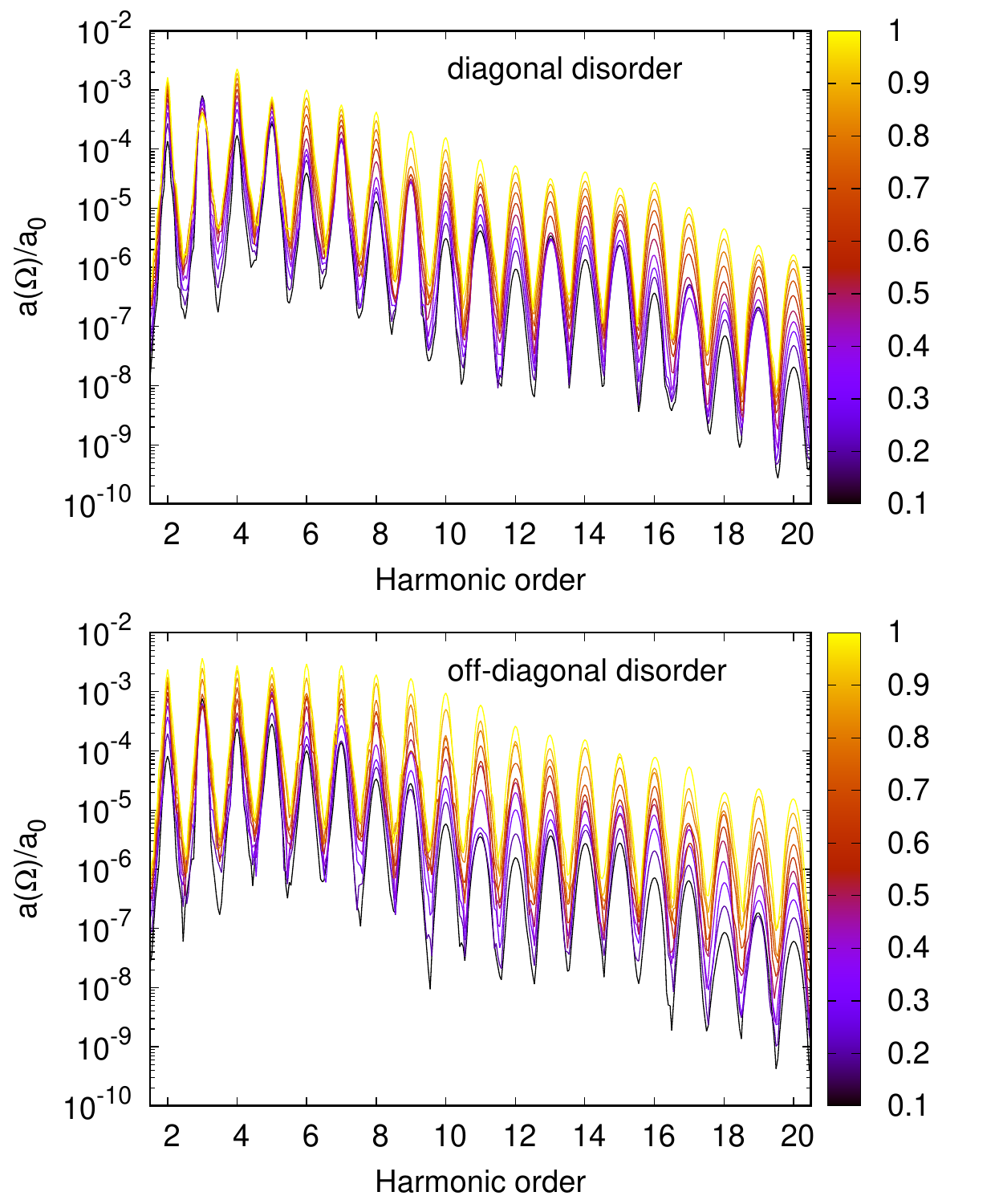}
\caption{Normalized HHG spectra for $C_{60}$ in logarithmic scale for a
range of the diagonal disorder (upper panel) and off-diagonal disorder
(lower panel). The color bar shows the strength of the disorder in eV. Laser
parameters same as those in Fig. \protect\ref{fig3}}
\label{fig4}
\end{figure}

\begin{figure}[tbp]
\includegraphics[width=0.39\textwidth]{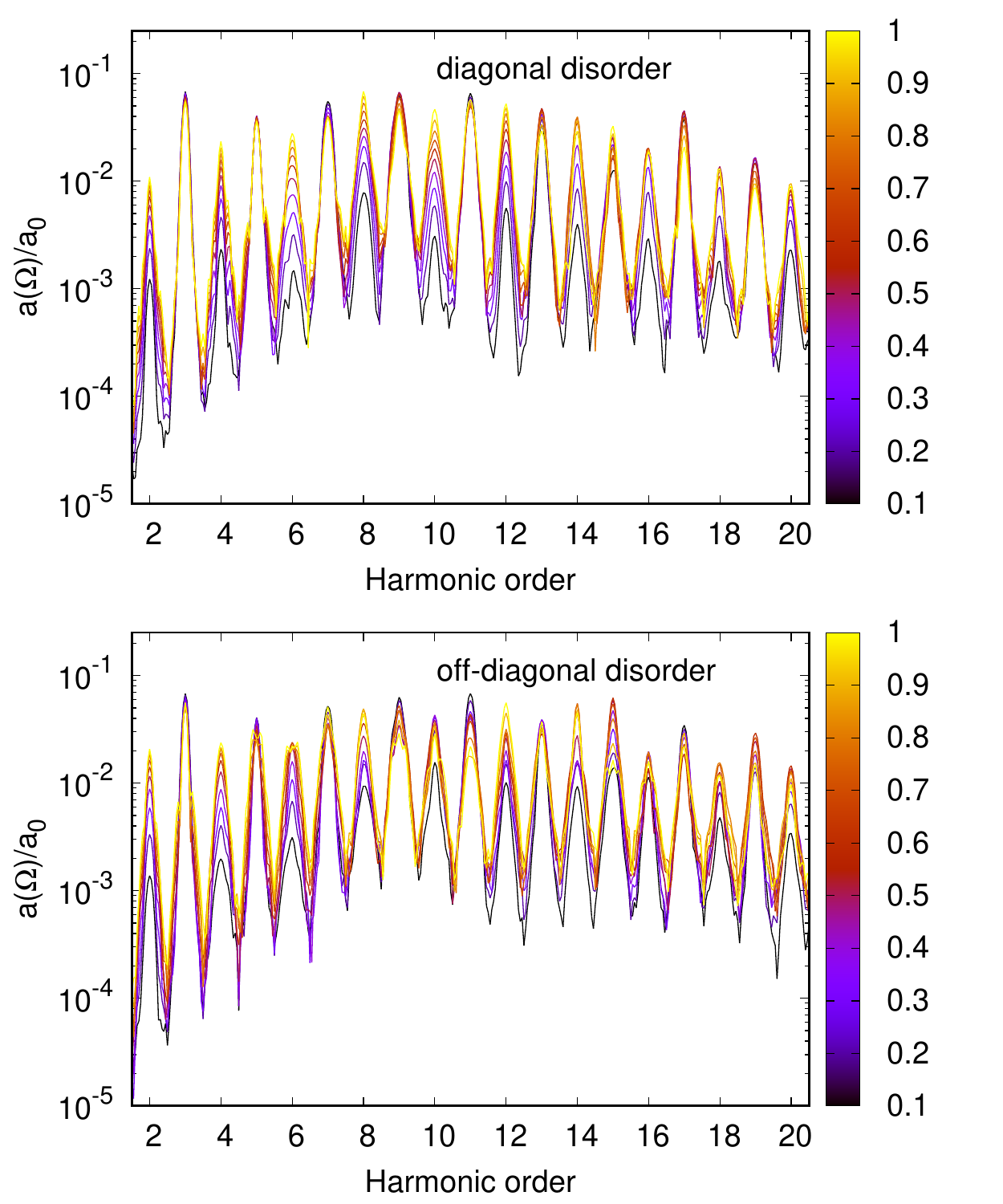}
\caption{The same as in Fig. \protect\ref{fig4}, but for $C_{180}$.}
\label{fig5}
\end{figure}

Next we study the dependence of high-harmonic emission on the disorder
strength. In Figs. \ref{fig4} and \ref{fig5} the HHG spectra up to the 20th
harmonics for $C_{60}$ and $C_{180}$ are shown for a range of the diagonal
and off-diagonal disorders $0.1-1.0$ eV. First we consider the disorder
effect on odd harmonics, which solely represent the HHG spectrum in the
disorder-free case. Here we observe a relative stable behavior of $C_{180}$
system changing the disorder strength. The harmonic yield increases slightly
with the increase of the diagonal disorder, $V_{\mathrm{on}}$, while the
effect of the off-diagonal one, $V_{\mathrm{off}}$, is not strongly
monotonic increasing. The influence of the disorder on the $C_{60}$ system
is much more pronounced. Overall we obtain an increase in odd-order
high-harmonics emission with increasing the disorder. Particularly, the
enhancement caused by the off-diagonal counterpart is by an order higher
compared to the diagonal one. Even-order harmonics, which appear due to the
breakdown of the symmetry caused by the disorder, increase with the disorder
strength. Figs. \ref{fig4} and \ref{fig5} show that the effect of the
disorder on the even-order harmonics is much more pronounced in the case of $%
C_{60}$ system compared to the $C_{180}$ one. The reason for the enhancement
of HHG emission lies on the fact that both, the diagonal and off-diagonal
disorders lift the degeneracy of states, and the parity of states becomes
uncertain opening up new numerous channels for interband transitions.
Surprisingly the disorder, being random in nature, does not initiate chaotic
behavior in the HHG spectra even for rather large disorder strengths. 

Examining Figs. 5 and 6, it becomes evident that the bandwidths of harmonics
remain almost resilient to varying disorder strengths. This phenomenon can
be attributed to the fundamental underpinnings of HHG within the
strong-field regime. In scenarios where perturbation theory is no longer
applicable due to the strength of the fields, the generation of high
harmonics is governed by sub-cycle nonlinear quantum dynamics, intricately
linked to the structure of the pump wave pulse. As the harmonic order
increases, harmonics emission primarily occur in proximity to the peak of
the pump wave, ultimately leading to the determination of harmonic
bandwidths by pulse duration and wave frequency. It is noteworthy that
higher relaxation rates can, in fact, result in a clearer HHG spectrum \cite%
{vampa2015semiclassical}.

Regarding the influence of disorder, as evident from Fig. 1, the presence of
disorder opens up novel channels for interband transitions, thereby
enhancing the likelihood of multiphoton electron excitation from the valence
to conduction bands. This process involves the creation of electron-hole
pairs followed by their subsequent recombination, resulting in the emission
of high-energy photons. To gain a quantitative understanding of this
phenomenon, we employ a Morlet transform of the dipole acceleration

\begin{equation}
\mathrm{a}\left( t,\Omega \right) \mathrm{=}\sqrt{\frac{\Omega }{\sigma }}%
\int_{0}^{\tau }\mathrm{dt}^{\prime }\mathrm{a}\left( t\right) \mathrm{e}%
^{i\Omega \left( t^{\prime }-t\right) }\mathrm{e}^{-\frac{\Omega ^{2}}{%
2\sigma ^{2}}\left( t^{\prime }-t\right) ^{2}}\mathrm{.}  \label{wavelet}
\end{equation}

For a concrete illustration, we consider the time profiles of the 3rd and
19th harmonics computed through Eq. (\ref{wavelet}), as depicted in Fig. \ref%
{TFP}. Irrespective of the presence of disorder, the qualitative behavior
remains consistent. For the 3rd harmonic (0.6 eV), which lies below the band
gap and is a product of intraband transitions, the time profile exhibits a
relatively smooth evolution over time, closely resembling the envelope of
the pump wave. Conversely, the 19th harmonic (3.8 eV), arising from
interband transitions, reveals a time profile characterized by two distinct
bursts within each optical cycle, corresponding to the maxima or minima of
the pump wave. Remarkably, these bursts occur in close proximity to the peak
of the pump wave.

\begin{figure}[tbp]
\includegraphics[width=0.45\textwidth]{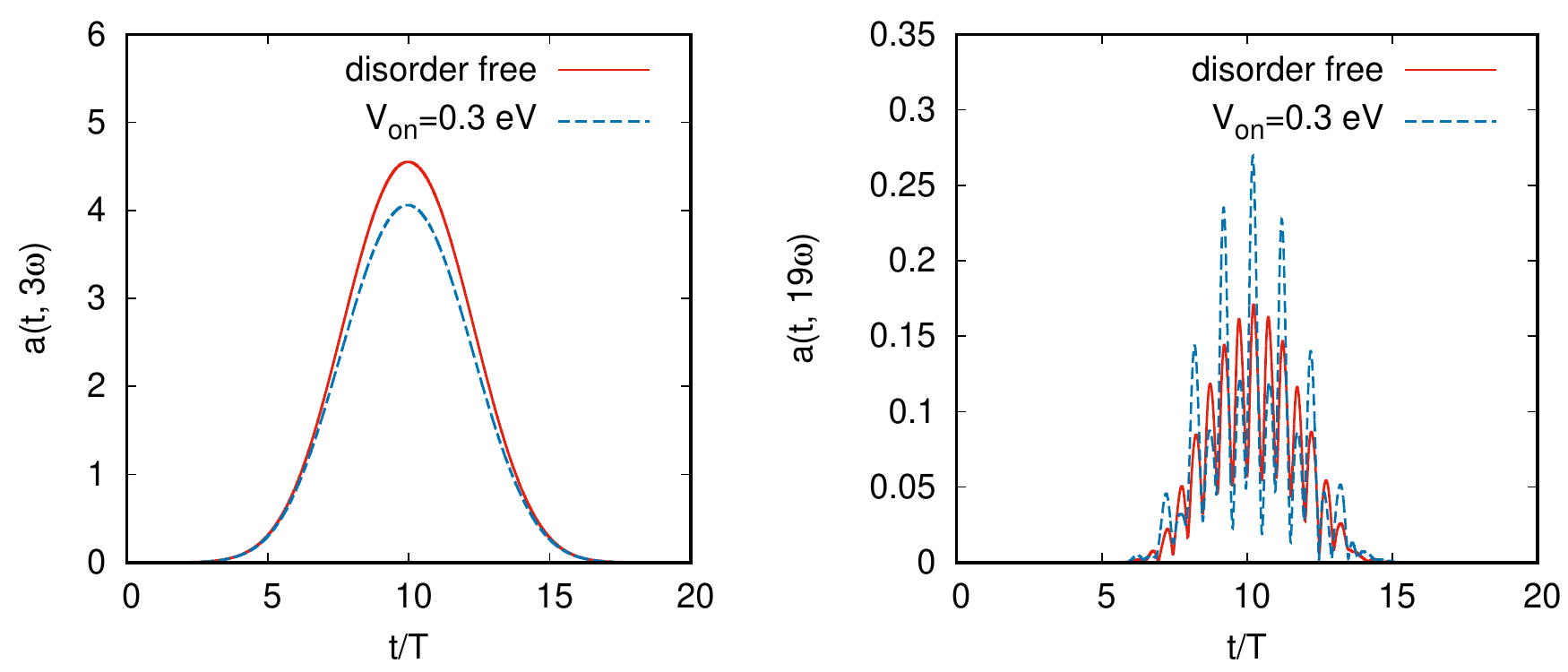}
\caption{The time profiles of the 3rd (left panel) and 19th (right panel)
harmonics for $C_{180}$. The interaction parameters are the same as those in
Fig. \protect\ref{fig4}. The Morlet transform parameter in Eq. (\protect\ref%
{wavelet}) is taken to be ${\protect\sigma }=15$}
\label{TFP}
\end{figure}

\begin{figure}[tbp]
\includegraphics[width=0.4\textwidth]{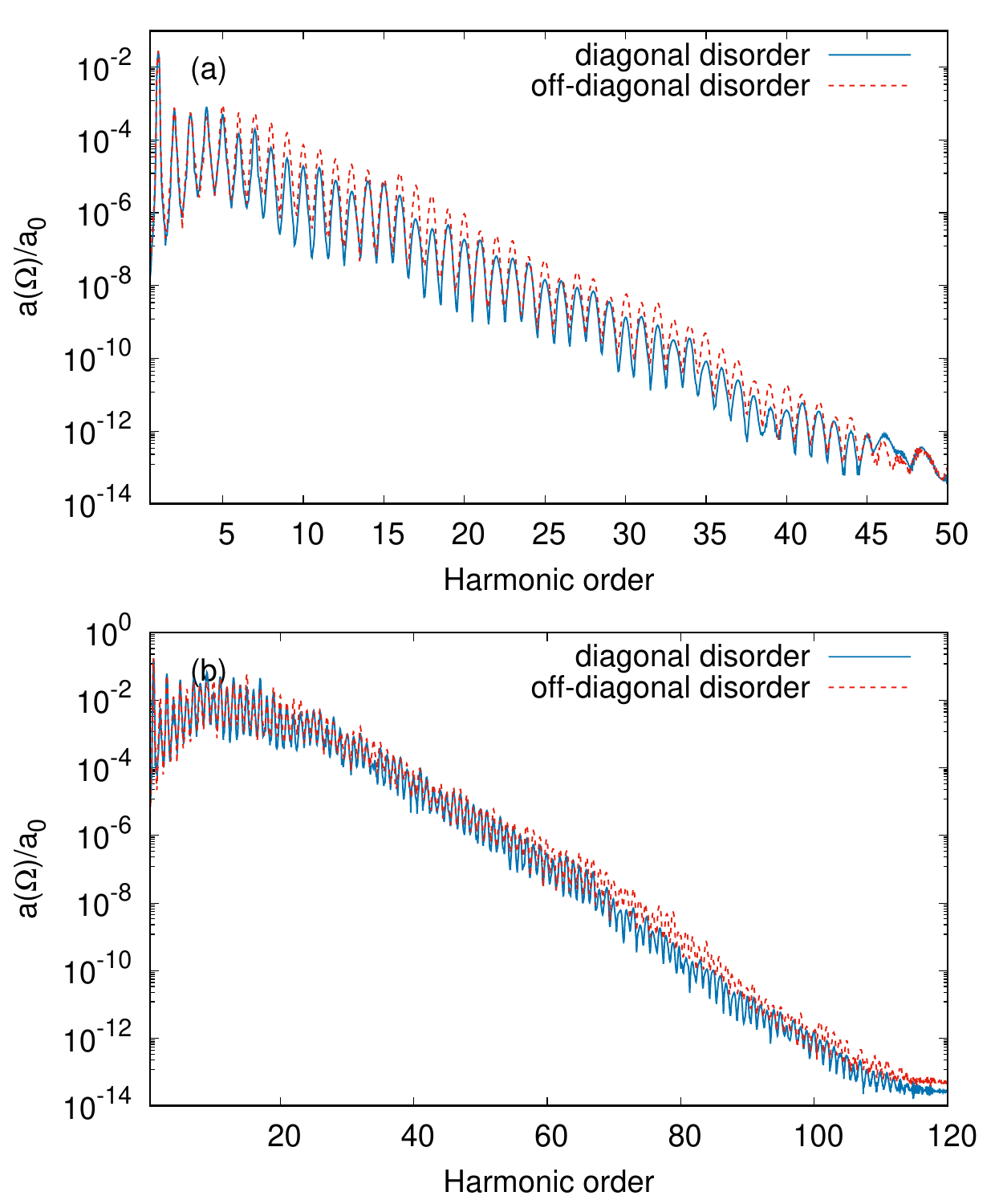}
\caption{Comparison of the influence of diagonal $V_{\mathrm{on}}=0.5\ 
\mathrm{eV}$ and off-diagonal $V_{\mathrm{off}}=0.5\ \mathrm{eV}$ disorders
on the HHG spectra for $C_{60}$ (a) and for $C_{180}$ (b). The wave
amplitude is taken to be $E_{0}=0.2\ \mathrm{V/\mathring{A}}$.}
\label{fig6}
\end{figure}

\begin{figure}[tbp]
\includegraphics[width=0.48\textwidth]{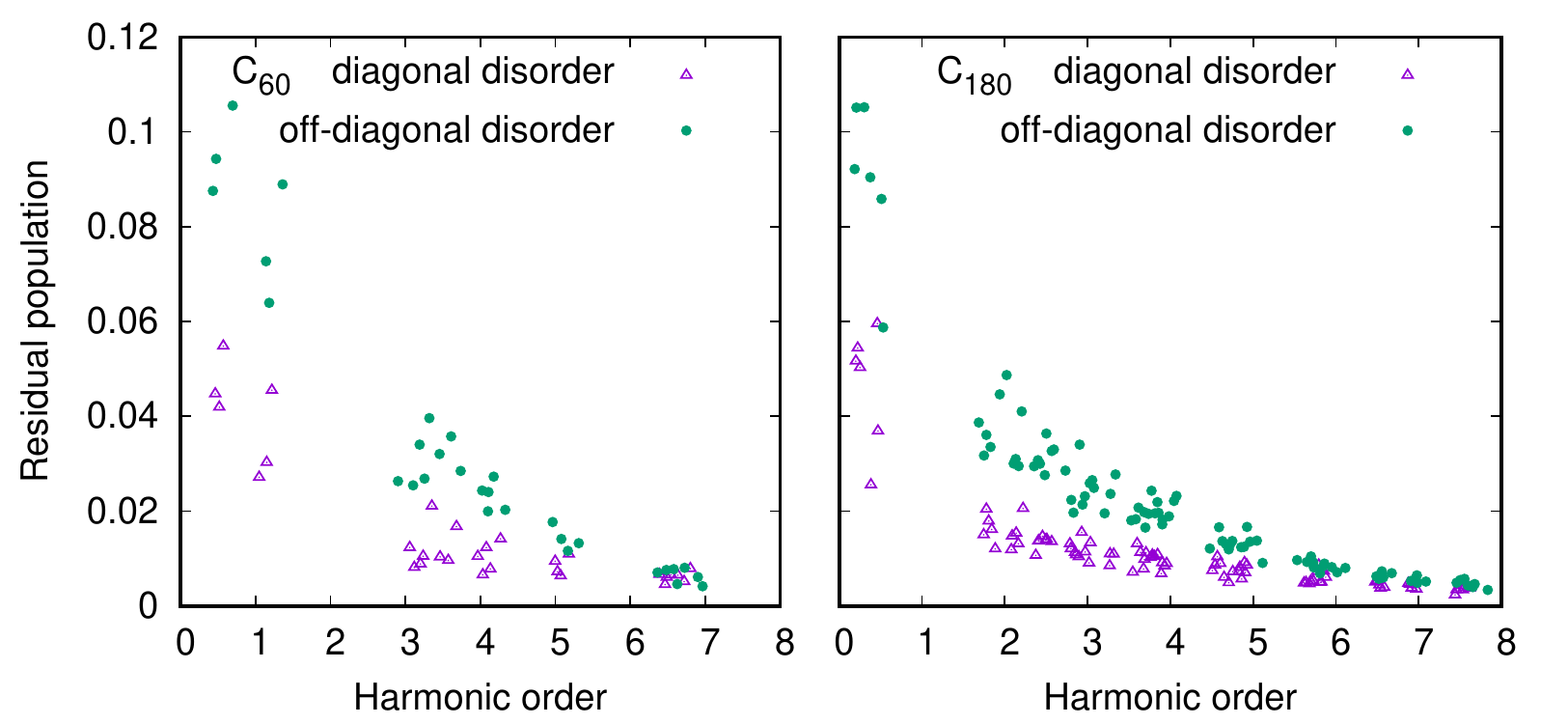}
\caption{The residual population of conduction band energy levels for the
setup of Fig. \protect\ref{fig6} for $C_{60}$ (a) and for $C_{180}$ (b).}
\label{fig7}
\end{figure}

Fig. \ref{fig6} compares the influences of diagonal and off-diagonal
disorders on the HHG spectra for a moderate disorder strength. From this
figure we observe, once more, stronger effect of the off-diagonal disorder
compared to the diagonal one in the amplification of the HHG yield. The
off-diagonal disorder acting on the hopping integral makes electrons more
mobile. Besides, the first dipole-allowed transition gaps are narrower for
the off-diagonal disorder, cf. Fig \ref{fig1}. As a result, the excitation
of electrons from the valence to the conduction band and further intraband
excitations to higher levels are more probable for the off-diagonal
disorder. In Fig. \ref{fig7} the residual population of conduction band
energy levels is displayed, which confirms the above-mentioned arguments. 

\begin{figure}[tbp]
\includegraphics[width=0.48\textwidth]{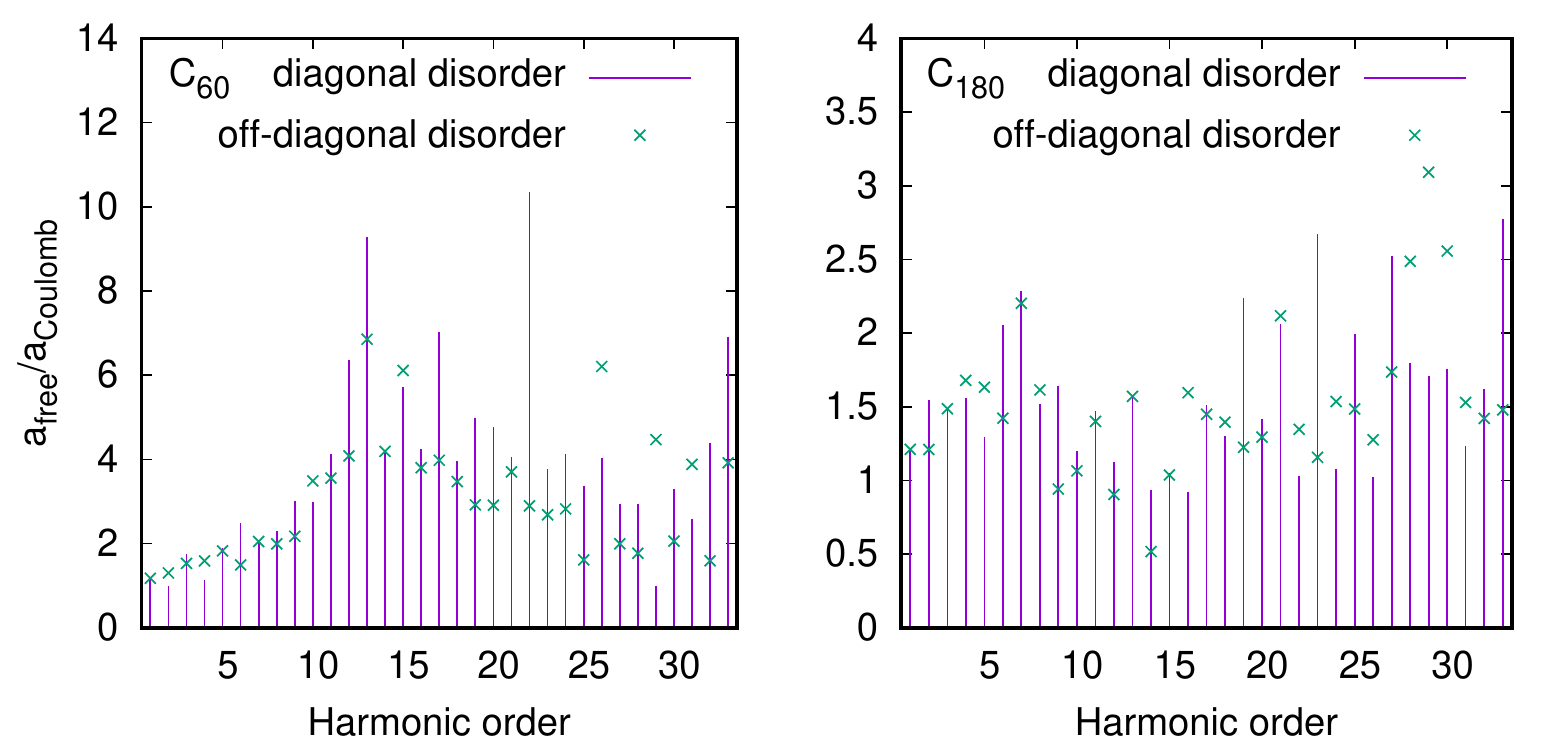}
\caption{Coulomb interaction effect on HHG via the ratio of the
high-harmonic intensities calculated without and with the Coulomb
interaction for $C_{60}$ (a) and for $C_{180}$ (b). For disorders strengths
we have taken $V_{\mathrm{on}}=0.5\ \mathrm{eV}$ and $V_{\mathrm{off}}=0.5\ 
\mathrm{eV}$. The wave amplitude is taken to be $E_{0}=0.2\ \mathrm{V/%
\mathring{A}}$.}
\label{fig8}
\end{figure}

In Ref. \cite{avetissian2021high} the suppression of HHG with the increase
of EEI has been found. Latter is connected with the fact that at strong
on-site and inter-site electron-electron repulsion the polarizability of
molecules, or in other words, electrons migration from the equilibrium
states, is suppressed. It is interesting to investigate the interplay of
Coulomb and disorder effects. In Fig. \ref{fig8}, the ratio of harmonic
intensities without and with the Coulomb interaction is shown. As we observe, 
in the case of the off-diagonal disorder the Coulomb-induced
suppression of HHG is smaller than in the case of the diagonal one. This is
connected with the fact that for the off-diagonal disorder the initial
configuration is almost homogeneous and close to the disorder-free case, cf.
Fig \ref{fig2}.

\begin{figure}[tbp]
\includegraphics[width=0.42\textwidth]{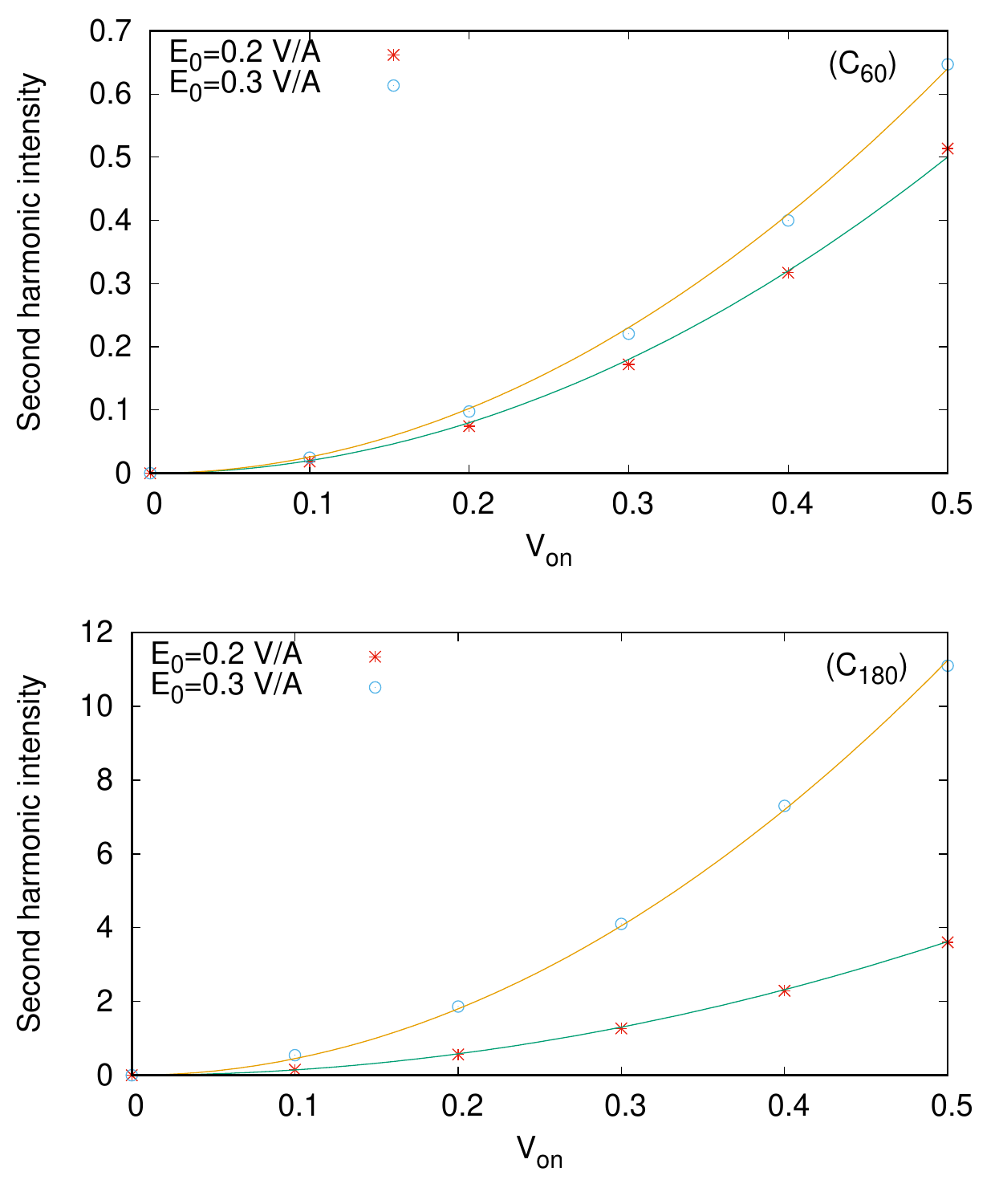}
\caption{The nonlinear response of $C_{60}$ and $C_{180}$ via the scaled
second harmonic intensity versus the strength of diagonal disorder. By solid
lines the fit of the second harmonic intensity by the law: $I_{2}\sim V_{%
\mathrm{on}}^{2}$ is shown.}
\label{fig9}
\end{figure}

Finally we turn to the second harmonic signal which is unique for diagnostic
tools. In Fig. \ref{fig9}, the nonlinear response of the fullerene
molecules, $C_{60}$ and $C_{180}$, via the scaled second harmonic intensity
depending on the diagonal disorder is displayed for different pump wave
intensities. As we see, the dependence of second harmonic intensity, $I_{2}$%
, on the diagonal disorder strength, $V_{\mathrm{on}}$, is perfectly
described by the law $I_{2}\sim V_{\mathrm{on}}^{2}$ for both, $C_{60}$ and $%
C_{180}$, systems. For the dependence of second harmonic intensity on the
off-diagonal disorder we have not observed such a universal law. Here we
report a faster increase of the intensity of the second harmonic with
respect to the increase of off-diagonal disorder strength, compared to the
diagonal one. In particular, at $E_{0}=0.2\ \mathrm{V/\mathring{A}}$ we have 
$I_{2}\sim V_{\mathrm{off}}^{3}$ and $I_{2}\sim V_{\mathrm{off}}^{5/2}$
dependencies, for $C_{60}$ and $C_{180}$ molecules, respectively.
Summarizing, our findings show that the non-linear optical response, in
particular the second harmonic signal, can be used as a spectroscopic tool
to measure the type and the strength of the disorder.

\section{Conclusion}

We have investigated the extreme nonlinear optical response of the
carbon-based quantum dots for the case of lattice topologies where sites are
distributed over a closed surface, subjected to different types of
disorders. In particular, we considered fullerene molecules $C_{60}$ and $%
C_{180}$ as typical examples of inversion symmetric stable configurations.
Solving the evolutionary equations for the single-particle density matrix,
taking into account the many-body Coulomb interaction in the Hartree-Fock
approximation, we demonstrate that the disorder-induced effects have strong
influence on the emission of high-harmonics. Both diagonal and off-diagonal
disorders lift the degeneracy of the states, opening up new channels for
interband transitions, leading to the amplification of high-harmonic
signals. For the characteristic odd harmonics the effect of disorder is
dependent on the type and the strength of the disorder, and on the molecular
system. We observe a drastic increase in the intensity of even-order
harmonics, making them comparable with the odd ones. The obtained results
show that the disorders have their unique footprints in the second harmonic
signal, which demonstrates a monotonic increase in intensity with the
different power laws for diagonal and off-diagonal disorders, which may
allow us to distinguish and measure the disorders type and strengths, paving
the way for an optical characterization of the disorder in nanostructures.

\begin{acknowledgments}
The work was supported by the Science Committee of Republic of
Armenia, project No. 20TTWS-1C010.
\end{acknowledgments}

\bibliographystyle{apsrev4-2}
\bibliography{bibliography}

\end{document}